# Persistent photoconductivity in 2-dimensional electron gases at different oxide interfaces

By *Emiliano Di Gennaro, Umberto Scotti di Uccio, Carmela Aruta, Claudia Cantoni, Alessandro Gadaleta, Andrew R. Lupini, Davide Maccariello, Daniele Marré, Ilaria Pallecchi, Domenico Paparo, Paolo Perna, Muhammad Riaz,* and *Fabio Miletto Granozio*

CNR-SPIN and Dipartimento di Fisica, Univ. di Napoli "Federico II",
Compl. Univ. di Monte S. Angelo, Via Cintia, I-80126 Napoli (Italy)
Materials Science and Technology Division, Oak Ridge National Laboratory,
1 Bethel Valley Road, Oak Ridge, TN 37831-6116 (USA)
CNR-SPIN and Dipartimento di Fisica, Univ. di Genova, via Dodecaneso 33, 16146, Genova (Italy)

**Keywords:** oxide-based interfaces, two dimensional electron gas, photoconductivity

We report on the transport characterization in dark and under light irradiation of three different interfaces: $LaAlO_3/SrTiO_3$, $LaGaO_3/SrTiO_3$, and the novel $NdGaO_3/SrTiO_3$ heterostructure. All of them share a perovskite structure, an insulating nature of the single building blocks, a polar/non-polar character and a critical thickness of four unit cells for the onset of conductivity. The interface structure and charge confinement in $NdGaO_3/SrTiO3$ are probed by atomic-scale-resolved electron energy loss spectroscopy showing that, similarly to $LaAlO_3/SrTiO_3$, extra electronic charge confined in a sheet of about 1.5 nm in thickness is present at the $NdGaO_3/SrTiO_3$ interface. Electric transport measurements performed in dark and under radiation show remarkable similarities and provide evidence that the persistent perturbation induced by light is an intrinsic peculiar property of the three investigated oxide-based polar/non-polar interfaces. Our work sets a framework for understanding the previous contrasting results found in literature about photoconductivity in $LaAlO_3/SrTiO_3$ and highlights the connection between the origin of persistent photoconductivity and the origin of conductivity itself. An improved understanding of the photo-induced metastable electron-hole pairs might allow to shed a direct light on the complex physics of this system and on the recently proposed perspectives of oxide interfaces for solar energy conversion.

## 1. Introduction

Oxide-based heterointerfaces can host physical properties that are absent in their individual bulk constituents. A striking example is the highly mobile two-dimensional electron gas (2DEG) generated at the interface between two band insulators, $LaAlO_3$ (LAO) and $SrTiO_3$ (STO).[1,2] Since the elements placed both at the A and B perovskites sites change their ionic valence state across the interface, a polar discontinuity is obtained. Ohtomo and Hwang indicated the polar discontinuity as a source of instability, resulting in an electronic reconstruction (ER) process and in the consequent formation of the interfacial 2DEG. [1,2] As proposed in[1] the ER mechanism is a displacement of electrons from the outer region of LAO into the Ti 3d states of the topmost STO layers driven by the relaxation of the electrostatic energy accumulated in the polar layer.[3,4] The ER scenario provides an elegant explanation of the characteristic phenomenology of the 2DEG, i.e., of the mentioned sensitivity to STO termination,[5] the observed depth of electron confinement,[6] and the minimum LAO thickness that is required to obtain the electron injection.[7] However, this model is still debated and different interpretations, pointing to the role of either cation intermixing at the interface,[8-11] or of oxygen vacancies introduced within STO during LAO growth,[12] were proposed. Presently, contrasting evidences were carried in support of either mechanism and no general agreement on the primary question regarding the origin of 2DEG has yet been found.

The transport properties of LAO/STO heterostructures are sensitive to light irradiation. This effect was first considered as detrimental for the realization of correct measurements of the intrinsic transport properties (see for instance Ref. [13-15]). More recently, the photoconductivity of LAO/STO attracted a considerable interest in view of potential optoelectronics applications and as a possible tool to investigate the fundamental physics of the system.[16-18] Seemingly contradictory data were reported so far about the threshold photon energy for LAO/STO photoconductance: room temperature measurements performed on metallic samples (thickness of LAO above 4 unit cell) showed a threshold for photoconductivity at about 380 nm,[14] corresponding to the indirect $SrTiO_3$ gap (about 3.3 eV);[19,20] while sensitivity to visible light (i.e., to photons with sub-gap energy) was reported for insulating, 3 unit cell thick LAO.[13,15] Finally, annealed samples showing a high resistance and a non metallic behavior showed a 5 order-of-magnitude drop in resistance under irradiation at 395 nm (about 3.1 eV, slightly below the indirect gap threshold).[18]

The $NdGaO_3/SrTiO_3$ (NGO/STO) epitaxial heterostructure shares with both LAO/STO,[1,21,22] and LGO/STO[23-26] some features considered as crucial for the (ER) to take place. The pseudocubic perovskitic cell of NGO has a low ($\approx 1\%$) mismatch with STO allowing for high quality epitaxial growth and thus for the fabrication of interfaces with high structural perfection. Similarly to LAO, NGO is constituted by a stack of positively charged A-planes and negatively charged B-planes, thus creating a polar discontinuity when epitaxially grown on STO. The dielectric constant of NGO ($\varepsilon_r \approx 20$) is also similar to the one of LAO,[27] suggesting within the ER model a similar critical thickness. Finally, the NGO gap is larger than that of STO (3.8 eV). This latter feature was proposed as an important parameter in polar/non-polar interface,[23] because it allows the conduction band of the polar layer to be presumably above the conduction band of STO, also according to the relative band alignment. Unlike LAO, NGO does not contain La, which was suspected to dope the interfacial STO, thus playing an important role in the origin of LAO/STO conductivity. [11,28,29] As reported in[30,31] and independently in[32], the NGO/STO interface exhibits metallic properties.

In this paper we compare the physical properties of the three mentioned interfaces. We aim at addressing a crucial question: is the transport behaviour of these three systems indeed similar, as expected on the basis of their common polar/non-polar character? Do they share the same mechanisms of photoresponse? However, before trying to answer, we first have to investigate NGO/STO properties, in consideration of the lack of information on this interface. The remaining part of the article is divided in two parts.

In the first part we demonstrate the similarities between NGO/STO, LGO/STO and LAO/STO. To this aim, we report on our investigation of the transport and structural properties of NGO/STO, based on the growth and characterization of about 50 samples.[30] We show that NGO/STO interfaces actually host a conductive 2DEG exhibiting a thickness threshold for conduction and sharing a very similar transport behavior with LGO/STO[23] and LAO/STO interfaces. We then discuss the dependence of such properties on the growth conditions, and in particular on deposition temperature and oxygen pressure. We show the microstructural characterization of the NGO/STO interfaces and analyze the distribution of $Ti^{3+}$ ions, concluding that an extra electronic charge is accumulated in proximity of the interface , as previously demonstrated for LAO/STO .

In the second part, once the similarity of LAO/STO, LGO/STO and NGO/STO in equilibrium conditions has been established, we address and compare the out-of-equilibrium electronic properties of our three type of interfaces by analyzing the transport during and after light irradiation. We focus on the different effect of above-gap and below gap irradiation, also analyzing the response in the time domain and discussing the extremely long response times of these systems. The results provide some clues on both the fundamental mechanisms common to NGO/STO, LGO/STO and LAO/STO, and on the recently proposed perspectives of band-engineered polar/non-polar interfaces for solar energy conversion. [33]

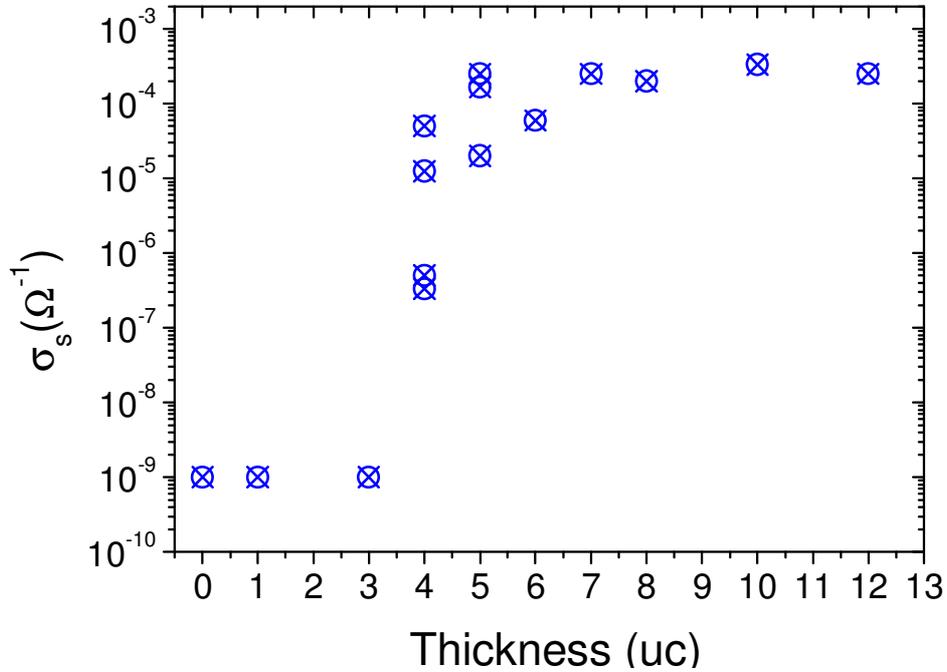

**Figure 1.** Dependence of the sheet conductivity on the number of NGO unit cells for NGO/STO samples fabricated at 700°C and $P(O_2)$ = 5·10$^{-2}$ mbar.

## 2. Experimental result and Data Analysis
### 2.1 - Transport properties of NGO/STO interfaces

**Figure 1** shows the room temperature sheet conductivities of a set of samples fabricated in the conditions indicated in the experimental section. NGO/STO interfaces were insulating when the NGO epitaxial layer has a thickness below 4 unit cells. Thicker samples were instead conducting. This threshold behavior is characteristic of both LAO/STO[7] and LGO/STO[23] and is generally considered as a typical feature of transport in polar-non polar interfaces. Interestingly, a major sample-to-sample variability of the room temperature conductance is found close to the critical thickness (4 and 5 u.c.), a behavior that we did not see in LAO/STO. Variable temperature transport measurements, including sheet resistance, magneto-resistance and Hall resistance were carried out on a batch of samples from 2.5 K to room temperature and in magnetic fields up to 9 T. The measurements were performed in standard Hall geometry resorting to ultrasonic bonding of the contacts. The R(T) curves of NGO/STO samples are qualitatively similar to those of LAO/STO and LGO/STO (**Figure 2a**). The R(T) curve of a sample annealed immediately after deposition for 1 hour at 700°C and 100 mbar $P(O_2)$ is also reported in Figure 2a. Similarly to the case of high pressure grown LAO/STO samples,[26] no effect on the conducting properties is seen. This observation confirms that NGO/STO is stable against oxygen post-annealing, similarly to LAO/STO[34] and LGO/STO.[26,27] NGO/STO samples grown in non optimal conditions, as mentioned before, are instead unstable against annealing.

Hall effect measurements[35] (Figure 2b) also demonstrate similarities of NGO/STO with the other cited interfaces, both in terms of carrier density and of electrical mobility. Typical values for NGO/STO are a room temperature sheet resistance of about of 10-20 kΩ, a carrier density, as extracted from the Hall resistance in a single band framework, of $10^{13}$-$10^{14}$ carriers cm$^{-2}$ and a related Hall mobility as high as 500 cm$^2$ s$^{-1}$ V$^{-1}$ at low temperature. Such values are both comparable to those for LAO/STO and LGO/STO interfaces prepared in the same setup,[23] and to previous determinations for NGO/STO.[32] A positive magneto-resistance is found at any field and temperature, being at the lowest temperature (T=2.5K) linear in $H^2$ up to 7 Tesla, while for higher fields it tends to saturate, with a deviation from linearity of 10% at 9 Tesla .

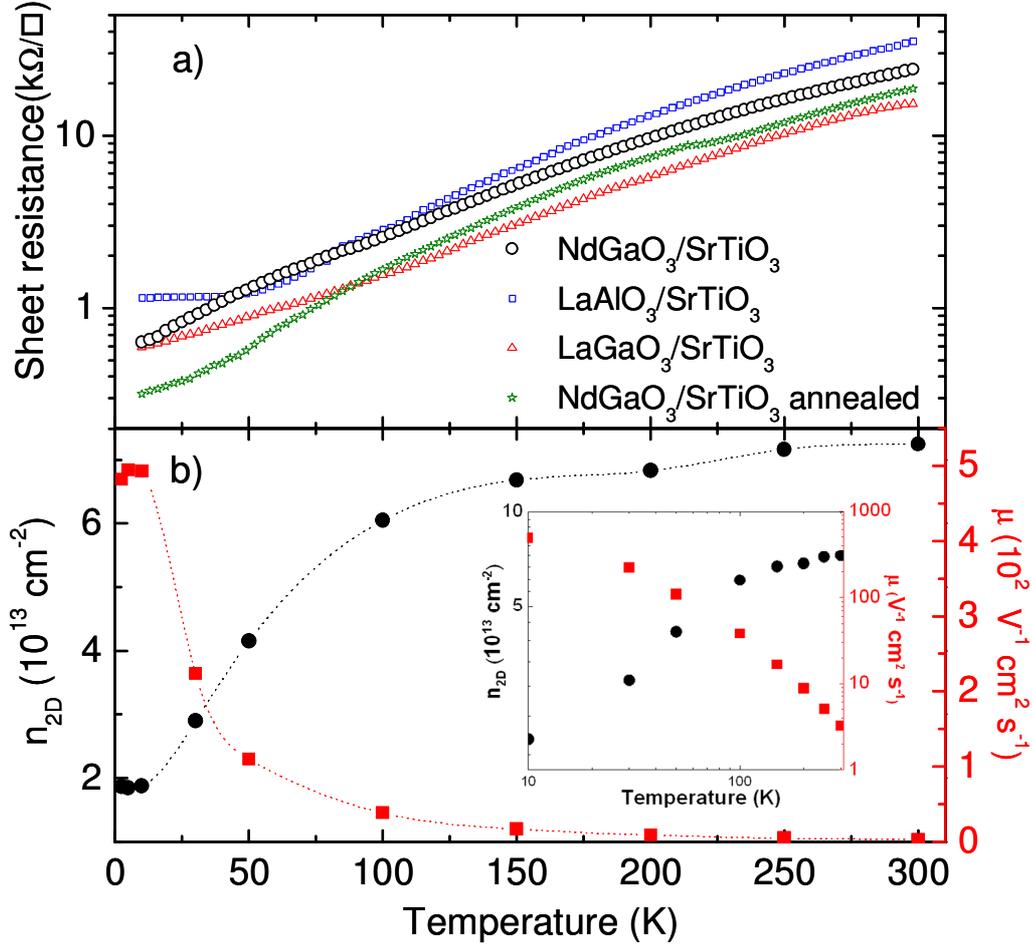

**Figure 2.** a): typical sheet resistance of NGO/STO, LGO/STO, LAO/STO vs. temperature; b) NGO/STO carrier density and mobility plotted as a function of temperature. A log-log plot is also shown in the inset.

## 2.2 - Nanoscale structure of NGO/STO interfaces

We already reported on the microstructural characterization of LAO/STO and LGO/STO interfaces in.[23,36] The atomic structure of different NGO/STO samples was investigated by aberration corrected scanning transmission electron microscopy (STEM) and electron energy loss spectroscopy (EELS) measurements performed in a Nion UltraSTEM operated at 100 kV and equipped with a third generation C3/C5 aberration corrector and an Enfina EEL spectrometer.[23,36] Cross section images of the samples were acquired at several locations using both bright field (BF) and high-angle annular dark field (HAADF) detectors.

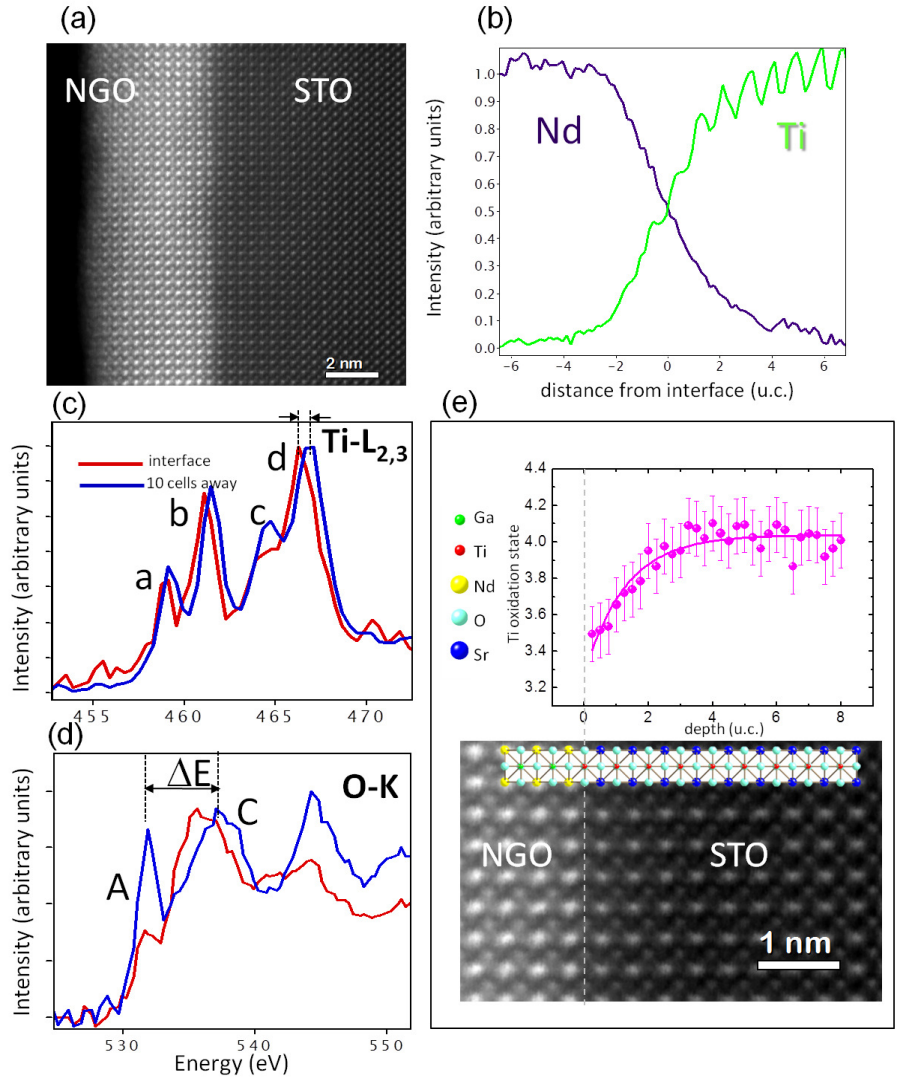

**Figure 3.** a) HAADF image of the interface b) Ti-$L_{2,3}$ and Nd-$M_{3,4}$ integrated profiles as function of distance from interface. c) and d) Ti-$L_{2,3}$ and O-K fine structure extracted from an EEL spectra acquired at the interface (red) and at 10 u.c. away from the interface (blue). e) Ti valence depth profile derived from the analysis of the O-K fine structure in the same spectrum image, and HAADF image of the NGO/STO interface with corresponding sketch of atomic stacking and same scale as the Ti valence plot.

In **Figure 3** we report the results of our TEM investigations. Figure 3a is a STEM image acquired with high angular annular dark field (HAADF) detector, showing a brighter contrast for the heaviest elements. All the images showed that the NGO films were free of structural defects and grew with a cube-on-cube epitaxial relationship of their pseudocubic cell on the STO cell. A slight cation intermixing was confirmed by repeated EELS scans across the interface and subsequent plots of the integrated intensities for the Ti-$L_{2,3}$ and Nd-$M_{4,5}$ edges and estimated to be 1.5 u.c., giving a predominant stacking sequence of the type SrO-$TiO_2$-NdO-$GaO_2$.

The spatial distribution of extra electronic charge at the STO interface was derived by analyzing changes in the fine structure of the EELS Ti-$L_{2,3}$ and O-K edges associated with a change of the Ti valence. Figures 3c and 3d show Ti-$L_{2,3}$ and O-K edges acquired at the interfacial $TiO_2$ plane and at 10 cells away from the interface in the STO bulk. The interfacial Ti-$L_{2,3}$ edge is shifted towards lower energy and peak c is significantly reduced with respect to the bulk edge. The O-K edge at the interface shows a decreased intensity for peak A and a reduced distance $\Delta E$ between A and C as compared to the bulk edge. These changes in the fine structures of Ti-$L_{2,3}$ and O-K edges are well documented in the literature and associated with a reduced valence of the Ti ion[37-39] resulting from charge transfer. Several methods exist for quantifying the Ti valence and here we used the O-K method,[40] consisting in acquiring reference spectra for STO ($Ti^{4+}$), LTO ($Ti^{3+}$), and TiO ($Ti^{2+}$) to derive a linear equation for the Ti valence as function of $\Delta E$ from which the Ti valence at the NGO/STO interface is extracted by interpolation, resulting in the graph shown in Figure 3e. The depth profile for the Ti valence follows an exponential law and indicates a confinement depth of ~ 3.5 unit cells. The total injected charge was obtained integrating along the depth profile the fraction of $Ti^{3+}$, yielding a value of 0.6 ± 0.2 e/unit cell area. This value is in agreement, within error, with the ER scenario. Note that part of the injected charge may be localized; therefore, this estimate cannot be directly compared with Hall measurements that only probe mobile carries.

**2.3 - Persistent photoconductivity of polar/non-polar interfaces**

The room temperature photoconductance measurements that are reported in the following were performed resorting to three light sources: a Hg lamp which emits several intense lines in the UV region, above the gap threshold E=3.3 eV, plus a visible component; a blue LED emitting at 455±20 nm (2.72 eV) radiation; a red LED emitting at 633±17 nm (1.96 eV) radiation. Such measurements have been chosen as a representative subset of a wider investigation based on several other light sources and performed also as a function of temperatures and magnetic field. In order to exclude trivial effects from the substrate, we preliminarily performed a series of test measurements on bare STO by depositing gold contacts on it. Under identical exposure conditions and with the same setup, the room temperature resistivity of such samples remained immeasurably high. In low temperature measurements, a feeble but measurable conductivity was found below 30K only under the light from the Hg lamp.

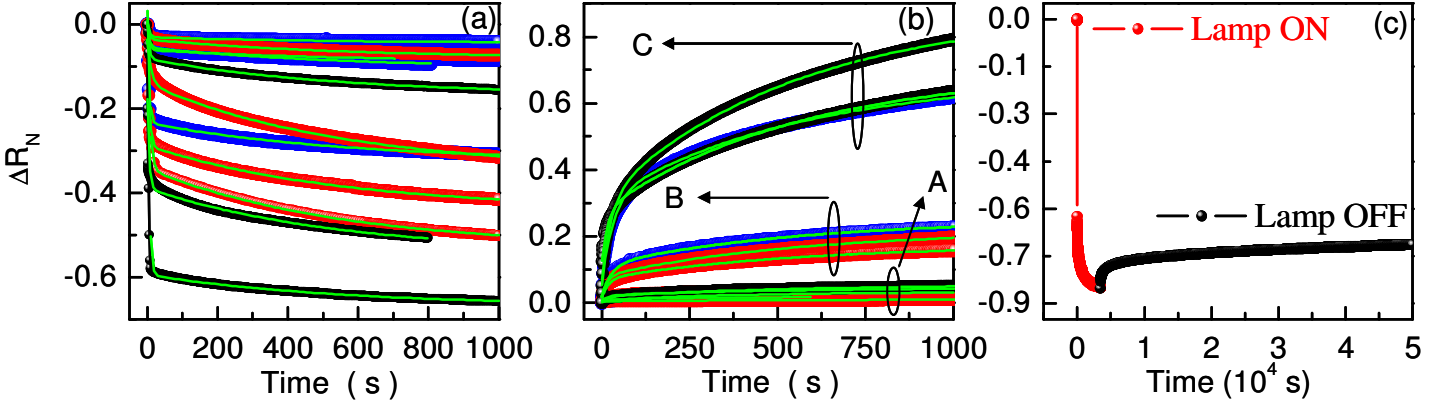

**Figure 4.** The behavior vs. time of the normalized photoinduced resistance variations is plotted vs. time. (a) Normalized resistance variation $\Delta R_N(t)$ for three NGO/STO interfaces (black), four LGO/STO interfaces (blue) and four LAO/STO interfaces (red) during UV light irradiation (light ON at t=0). The green lines are fits according to eq. (1). (b) Normalized resistance $\Delta R_N(t)$ for the same samples after UV irradiation (light OFF at t=0). The samples are divided in three groups according to their room temperature resistance. The green lines are again fits according to eq. (1). (c) Normalized resistance $\Delta R_N(t)$ decay and subsequent recovery collected on the NGO/STO sample belonging to the group C on a timescale of the order of half a day.

R(t) profiles under and after illumination by different light sources were collected for several samples of the three different interfaces with different room temperature sheet resistances, also as a function of temperature. The typical acquisition times of the R(t) profiles was $10^3$s, while longer delays (up to $10^4$-$10^5$ s) where explored occasionally. As a first comparison of the response of NGO/STO, LGO/STO and LAO/STO samples to light, we report in **Figure 4a** and **b** the room temperature, time dependent response of 11 different samples (three NGO/STO, four LGO/STO and four LAO/STO) under the UV light of the Hg lamp. Samples with high or very high room temperature resistance (typically NGO grown outside the correct parameter range, as defined above) were included on purpose within this set, in order to allow a direct comparison also with data reported in[18] and try to rationalize the contrasting previous results. The $\Delta R_N(t) = [R(t)-R(t=0)]/R(t=0)$ profiles registered during and after UV illumination show that a persistent photoconductivity effect is present in all samples. In spite of the marked sample-to-sample variability in the intensity of the effect, the data seem to follow a universal low that is independent of the sample, of the material (LAO, LGO or NGO) and also of the temperature. Qualitatively, the main feature is the existence of a relatively fast process, with timescale of the order of about 10 s, followed by a much slower tail. In principle, several functional expressions may be adopted to describe the latter, including logarithmic or stretched exponential functions. Both functions provide good fits of very long records of data collected in very long time ranges ($10^4$-$10^5$ s). In the following, however, we will present fits to the data in Figures 4a and b based on a conceptually much simpler double-exponential function

$$\frac{\Delta R(t)}{R_{t=0}} = A_1^{L,D}\left(1-\exp\left(-\frac{t}{\tau_1^{L,D}}\right)\right) + A_2^{L,D}\left(1-\exp\left(-\frac{t}{\tau_2^{L,D}}\right)\right) \quad (1)$$

where the indices L or D in $A_{1,2}^{L,D}$ and $\tau_{1,2}^{L,D}$ stand for light (Figure 4a) and dark (Figure 4b) respectively. The function reported in (1) has a number of advantages: it fits very successfully all the registered data on $10^3$s timescales adopted for the collection of our data; it allows for the identification of well defined time constants; it emphasizes in a very clear way, as shown below, the different behavior of the interfaces under light of different wavelength. The fits are reproduced in Figures 4a,b as solid green curves. During the global fit procedure shown in Figure 5a,b, the $\tau_{2,Hg}^{L}$ and $\tau_{2,Hg}^{D}$ values of all samples where kept equal (if left free to vary independently, their values would spread within ± 10% from the global fit value, with no correlation with the kind of interface). The $\tau_1^{L,D}$ values where left free to vary from sample to sample, as well as the $A_{1,2}^{L,D}$ values (the A coefficients being negative for the L case and positive for the D case).[41] The best fit values are: $\tau_{2,Hg}^{L} = 0.50 \times 10^3$ s, $\tau_{2,Hg}^{D} = 0.45 \times 10^3$ s and, depending on sample, 1s < $\tau_{1,Hg}^{L}$ < 10s ; 7s < $\tau_{1,Hg}^{D}$ <25s. In Figure 4b, the samples are divided in three groups A, B and C having respectively a room temperature sheet resistance $R^{D}_{300\,K}$ in the 1.5-4x10$^4$Ω range, in the 1-3x10$^5$ Ω range and in the 10$^6$ Ω range. This allows to highlight a correlation between the intensity of the photoresponse (i.e. the modulus of the A coefficients) and $R^{D}_{300\,K}$.

Highly conductive samples with room temperature sheet resistances of the order of 10-20kΩ show small relative resistance variations, of the order of a few percent for UV light at the fluences employed in this work. A very intense photoresponse, is found instead in oxide-based interfaces presenting a high resistivity or even a semiconducting behavior in dark, as shown in[16] for the case of LAO/STO. This is clearly shown in the data reported in Figure 4c, collected on the NGO/STO sample belonging to the group C. Figure 4c also shows that the time scale for a full recovery of the pristine resistance value is very long compared to the time duration of typical laboratory experiments. In fact, 5x10$^4$ s after the end of the illumination only about 20% of the resistance loss occurred under radiation is recovered.

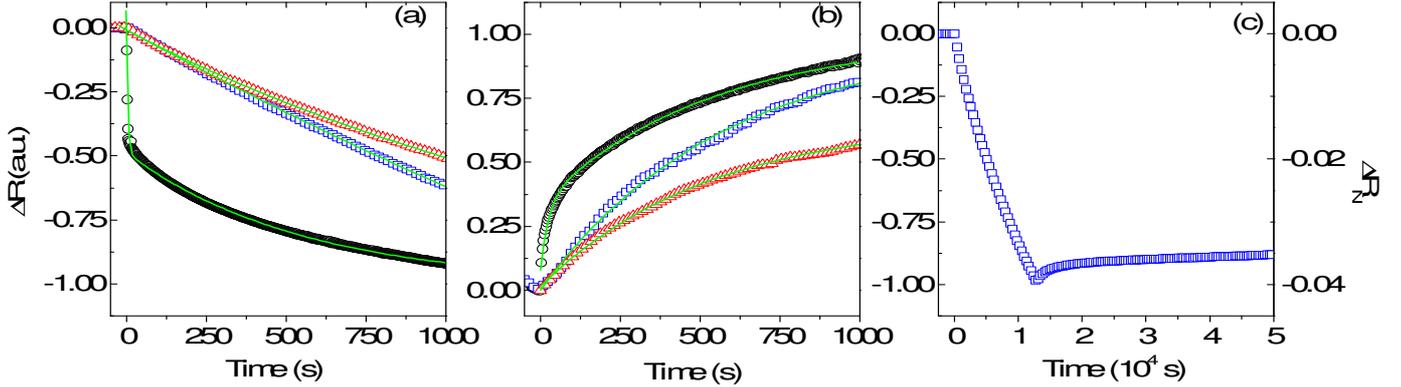

**Figure 5.** The temporal profile of the resistance variation of the NGO/STO sample belonging to the A group (as defined in fig. 4b), during and after the irradiation by UV (black circles), blue (blue squares) and red (red triangles) light. (a) Temporal profile of resistance variation ΔR under irradiation. (b) Temporal profile of the resistance variation ΔR after irradiation. (c) Resistance decay of the same NGO/STO sample under blue light and subsequent recovery, collected on a timescale comparable to Figure 4c. The left scale of fig. 5c is the same as fig 5a and b, while the right scale allows direct comparison with fig. 4c.

**Figure 5** compares the photoconductivity dynamics of the NGO/STO sample belonging to set A, collected under and after irradiation by UV, blue and red light. The photoresponse intensity qualitatively decreases with increasing wavelength. Nevertheless simply plotting the ΔR$_N$(t) data for the three sources on the same graph would be misleading, since the amplitude of the variation changes by changing the intensity of each source. A quantitative and physically significant analysis of the absolute intensities ($A^{L,D}_{1,2}$ coefficients) would require extra information (e.g. an absolute calibration of the fluences of the different sources, including the full spectral information of the Hg lamp; the different absorption coefficients at various wavelengths; the different penetration depths), but would not be relevant to the physics addressed in this work. We concentrate instead on the more important observation that when equation (1) is applied to fit the dynamics under visible light, the short timescale contributions (i.e. the $A^{L,D}_1$ coefficients) vanish. To make this evident the ΔR values in Figure 5a,b are normalized arbitrarily, with the only aim of plotting the data on the same scale. Only $\tau^{L,D}_2$ values are therefore here specified, their values being: $\tau^{L}_{2,red}$ = 1.8x10$^3$ s, $\tau^{L}_{2,blue}$ = 4.5x10$^3$ s, $\tau^{D}_{2,red}$ = 0.46x10$^3$ s, $\tau^{D}_{2,blue}$ = 0.55x10$^3$ s. Quite interestingly the time constants $\tau^{D}_{2,red}$ and $\tau^{D}_{2,blue}$ are relatively close to the $\tau^{D}_{2,Hg}$ value 0.45x10$^3$ s, obtained from the global fit in Figure 4b. The effect of the blue light on a longer timescale is reported in Figure 5c. By comparing the photoresponse dynamics in the two quite different cases reported in Figure 4c and 5c (a highly resistive sample under UV light and a standard sample under visible light) we find both similarities and differences. The similarities lie in the extraordinarily long persistency of the effect, that 5x10$^4$ s after the end of irradiation is still very far from a full recovery; the differences are found in the intensity of the effect (about an order of magnitude in Figure 4c, a few percent in 5c) and in the presence of a relatively fast component of the recovery only under UV light.

**3. Discussion of the mechanisms determining the persistent photoresponse**

We found that, beside possessing in principle all the key ingredients for an ER system, NGO/STO interfaces share with LAO/STO and LGO/STO (within sample-to-sample variability) the same transport properties, the same critical thickness and the capability (unlike amorphous oxide-based interfaces) [42] to be grown conductive at high oxygen pressures and to survive an oxygen post-annealing. The electron gas in NGO/STO is confined at the interface, as demonstrated by the EELS data on Ti$^{3+}$ distribution reported in Figures 3c,d and e. Both the depth of the electron gas and the total charge density are in good agreement with LAO/STO[36] and with the expectations of the ER model. The data in Figure 4 suggest that the quite unique response to light of different kinds of polar/non-polar interfaces is essentially identical within sample-to-sample variability and must be attributed, on the base of its dynamics, to the same mechanism(s).

A first insight on the role that the polar layer plays in modifying the electronic level structure at the STO interface is provided by the comparison of interfaces response with the case of bare STO. While the absence of photoresponse in STO under sub-gap irradiation is obvious, the insulating behavior under irradiation by above-gap photons from the Hg lamp requires a quantitative evaluation. A surface electron density of $10^{12}$ cm$^{-2}$ was reported to be necessary to induce an insulator to metal transition at the STO surface.[43] The upper limit for the equilibrium population $n_{eq}$ of photocarriers (corresponding to a 100% photocarrier generation yield) equals the areal flux F of incoming above-gap photons times the average exciton lifetime $\tau$ ($n_{eq}=F\tau$), which was measured to be of the order of 2-3 ns in STO single crystals.[44,45] An incident flux of above-gap photons exceeding $10^{20}$ cm$^{-2}$s$^{-1}$ (i.e., a fluence F > 10W/cm$^2$, orders of magnitude above the fluence of our Hg lamp) would be required to sustain a population of $10^{12}$ cm$^{-2}$ carriers. This simple argument shows that, by converse, the origin of interface photoconductivity must lies in a giant increase (exceeding a factor $10^9$) of STO carriers lifetime induced by the presence of the polar layer.

The persistent photoconductivity in semiconductor-based interfaces hosting a 2DEG , is interpreted in terms of a separation in real space of the electron-hole pairs by an intense built-in electric field.[46-49] Such separation hinders the recombination process and leads to very long lifetimes of the photoexcited states. We will adopt similar arguments to discuss the properties of oxide interfaces.

Let's start by modeling the carriers populations in the interface metallic ground state. Whatever mechanism we assume for interface conductivity, the charge neutrality requires that the electrons moving to the 2DEG leave behind an equal number of holes, which are trapped at specific sites and do not contribute to the conduction. The ground state can therefore be modeled as being populated by stable electron–hole pairs, where the mobile electrons populate a narrow quantum well at the interface at the STO side, as known for LAO/STO and experimentally confirmed for NGO/STO by the EELS measurements in this work.

In this picture, the intrinsic and extrinsic explanations for interface conductivity differ in the fact that they attribute different positions to the donors states where the holes reside: within the ER model, such donors states are believed to lie at the surface of the polar layer; the alternative mechanisms based on the presence of oxygen vacancies or cation intermixing assume that they lie in STO. When the ground state is perturbed by light, extra excitation in the form of photoelectron-photohole pairs are formed and add to the stable pairs of the ground state. Some of these photoinduced excitations are characterized by an extraordinary long lifetime, i.e. they are metastable. Electrons are transferred to the interface 2DEG, as confirmed by the fact that their conductivity adds to the conductivity of the unperturbed interface. As in the case of the ground state, it is the position of the donor states where the photoinduced holes reside that has yet to be identified. Obviously, it is tempting to assume that the sites where of "stable" and "photoinduced" holes reside are the same. The decay dynamics of photoconductivity is determined by the lifetime of the photoinduced metastable excitations, i.e., by the recombination probability of the photoinduced electron-hole pairs. In turn, such recombination cross-section is determined by the position and by the electronic nature of the donor states hosting the photoinduced holes.[50]

The band diagram in **Figure 6** provides a sketch of the states available in proximity of the interfaces. The sketch takes into account both the band bending in the polar layer and in STO, as foreseen by the ER model (in agreement with theoretical and experimental reports regarding LAO/STO, such as ab-initio computations,[51] tunneling measurements[52,53] and polarization profiles mapped across the interface) [36] and the presence of localized states near the interface that are expected by the presence of point defects. Direct promotion from the VB to the CB is made possible under radiation containing above-gap UV components, as our Hg lamp. All the potential mechanisms that can be hypothesized to allow the CB occupation under radiation by visible light are instead strictly related to the electronic structure of the interface. In particular: the *(a)* mechanism is made possible by the band bending present at the STO side, and corresponding to the quantum well that confines electrons at the interface: such negative energy misalignment of STO $t_{2g}$ interface states with respect to bulk was estimated to be of the order of a few tenths of eV[54] and might possibly be sufficient to allow promotion in the conduction band by blue-violet photons; the *(b)* mechanism is related to the possible presence of point defects causing the formation of localized in-gap states on the STO side; the *(c)* mechanism is made possible by the fact that the polar layers valence band is bent upwards by the dipole electric field (an effect possibly mitigated, with respect to the sketch, by the presence of surface adsorbates[55] or of defect states inside the LAO gap acting as donors [56]).

It is straightforward to associate the fast component of the photoresponse (with $A_1^L$ coefficients and $\tau_1^L$ time constants) of equation (1), absent under the illumination by visible light, to the massive occupation of the STO conduction band (CB) by electrons directly excited from the STO valence band (VB). The interface is not necessary to allow this transition; however, as mentioned above, it increases the excitation lifetime, resulting in a higher non-equilibrium carrier concentration. The sketch in Figure 6 suggests that such increased lifetime (estimated by $\tau_1^D$) may be due to the space separation of the electron, which is pushed by the electric field towards the interface, and the hole, that is either pushed away from the interface on the STO or polar layer side or trapped in a localized state. It is reasonable that only a fraction of the photoinduced electron-hole pairs created very close to the interfaces are separated, while the majority of them (as in the case of bare STO) decay on the timescale of a few ns[45] and are therefore not detected in our transport measurements.

The decay times after irradiation by blue and red LEDs ($\tau_{2,red}^D$, $\tau_{2,blue}^D$) are both comparable with the slow components of the dynamics after irradiation by the Hg lamp. Such similarity suggests that the photoexcited states obtained after irradiation by red light, blue light and, after a transient, by Hg lamp light, are similar. We note that in any case the recovery is characterized by a long tail, so that the relaxation time is longer than the excitation time. This evidence is a further indication of the dragging effect of the electric field. During the excitation, the rise time of the photoconductance is the time required for the system to reach a stationary state, in which the on-site recombination prevails and balances the creation rate. In the meanwhile, the electric field displaces in opposite directions some e- and h+ before they recombine. When the radiation is turned off, part of the excitations is spatially separated; this will lead to a long tail in the relaxation process.

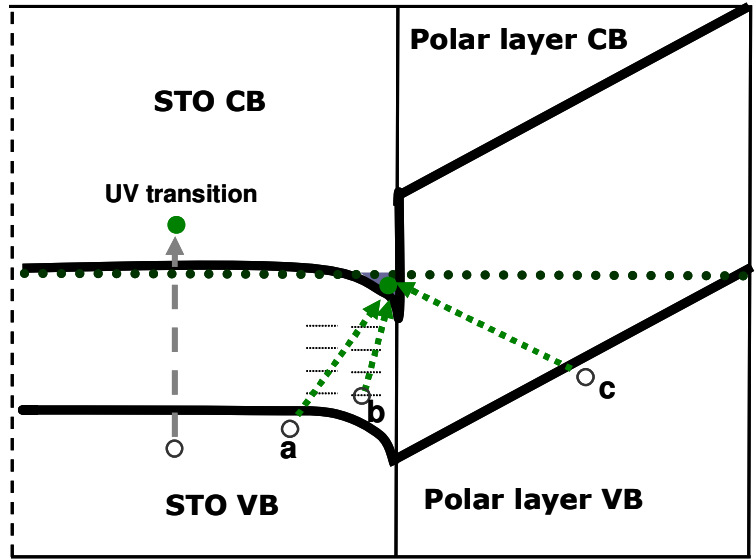

**Figure 6**. Sketch of the band structure of a polar/non-polar interfaces where the band bending is depicted as expected within an ER scenario and localized states possibly due to oxygen vacancies or intermixing are also considered in the gap. The bending of the STO band is emphasized for clarity. The possible mechanisms (a), (b) and (c) described in the text for the promotion of carriers at the interface into the STO CB by sub-gap photons are sketched. A standard inter-band transition induced by above-gap UV light is also shown

We can now discuss the nature of the excited states in terms of their single elementary excitations, that is, electron-hole pairs. The *(a)* mechanism cannot justify the photoconductivity effect by red light; we discard it since we are looking for a single mechanism valid at all the measured wavelengths. The *(b)* process corresponds to the photoexcitation of defect states lying at the interface, as proposed in ref. [18]. This mechanism may in principle be effective in a wide range of wavelengths and may possibly justify the extended carrier pair lifetime, especially if donor states are available in STO far enough from the interface, since this would reduce their wavefunction overlap with the electrons confined in the 2DEG. Let's now discuss as an alternative explanation the *(c)* mechanism, that can be seen as a photoinduced ER and appears of greater interest. In this photoexcited states the two photocarriers are physically separated at the two sides of the interface. This is the exact analogous to the above mentioned cases of persistent photoconductivity in semiconducting 2DEGs systems[46-49]. It is also analogous to the charge separation effect in a photovoltaic p-n junction and it would correspond to the emergence of the photovoltaic effects recently proposed in LaAlO$_3$/SrTiO$_3$[33,57] and more generally in oxide heterostructures.[58] Such mechanism gains therefore a great potential interest and could be directly probed by applying a transparent electrode above the polar layer and measuring a photocurrent.

The above reported analysis obviously applies to subcritical thick samples, as the ones reported in [13],[14] and to samples showing an insulating behavior, as those reported in [18]. The photoinduced carriers, that under visible light provide a negligible contribution to conductivity in the case of samples that are already metallic in dark, dominate the conduction in insulating samples. This was fully confirmed on our NGO/STO samples grown at too high temperatures, that switched to a metallic behavior under and after irradiation by visible light.

As a final observation, we remark that a system being so easily driven to an excited metastable state by light can be hardly characterized in its ground state by any spectroscopic technique using visible light, or smaller wavelength radiation, as a probe. In particular, long lived electron-hole couples of the *(c)* type would alter persistently the charge balance on the two sides of the interface, thus screening the built-in electric field and flattening the band bending. This is reminiscent of the surface photovoltage that is known to flatten under light the band bending at semiconducting surfaces[59] and interfaces.[60]

**4. Conclusions**

In this work we compared three different oxide interfaces based on the polar discontinuity concept: NGO/STO, LGO/STO and LAO/STO, with the aim of establishing to what extent is the physics of these three systems similar. We found that the optimal conditions for the growth of stable metallic NGO/STO interfaces are at higher oxygen pressures and lower deposition temperatures with respect to the typical conditons for LAO/STO. The conducting layer in NGO/STO is confined within few u.c. from the interface, as demonstrated by the Ti$^{3+}$ EELS profile, similarly to the case of LAO/STO. Furthermore, NGO/STO interfaces share with LAO/STO and LGO/STO the same transport properties, the same critical thickness and the same stability against oxygen post-annealing.

The photoresponse was employed as a tool to compare the electronic properties of NGO/STO with LGO/STO and LAO/STO. We found a strong room temperature photoconductivity effect in all of the three STO-based interfaces, differently from the case of bare STO where a feeble effect in only visible in our measurement conditions below 30K. The transport properties of all the analyzed interfaces exhibit a very specific and peculiar response to light, with a persistent photoconductivity effect showing a universal dynamics whose intensity depends on the room temperature sheet resistance of the sample rather than on the nature of the polar layer. We quantitatively analysed the time dynamics and found remarkably long timescales and a dramatic dependence of the photoresponse on the wavelength. This dependence is not only of quantitative but also of qualitative type, since the functional form vs. time changes when the photon energy exceeds the STO gap threshold. We attribute this to the emergence of different photoinduced excited states with different decay dynamics under irradiation by above-gap and below-gap photons. The possible photoinduced metastable excitations created at the interface by either UV or visible light are discussed in terms of a typical polar/non-polar interface band structure, also taking into account the possible presence of point defects. Such approach allows us to identify and discuss the nature of the long-lived electron-hole pairs that are potentially responsible for the peculiar ultra-slow dynamics of the observed photoresponse. Our analysis is in agreement with the recently proposed analogy between oxide interfaces and photovoltaic junctions. The persistent perturbation induced by light suggests that photon-based spectroscopies, when applied to this system, might probe samples significantly perturbed from their ground state. We argue that persistent photoconductivity is related to the presence of strong interface electric fields and that the origin of photocarriers might be intimately related to the origin of the electrons already forming, before illumination, the 2DEG.

## 5. Experimental section

*Film Growth* - NGO films, as well as LGO and LAO films, were deposited by Reflection High Energy Electron Diffraction (RHEED) assisted pulsed laser deposition (KrF excimer laser, 248 nm). The (001) STO substrates were chemically treated in de-ionized water, buffered-HF prior to deposition, in order to obtain a single $TiO_2$ terminating layer, and finally annealed at 950°C in order to get flatter step edges. An extensive growth optimization was performed by varying in a suitable interval the deposition temperature (650-800°C), the deposition pressure ($10^{-4}$-$10^{-1}$ mbar), the target to substrate distance (35-45mm) and the laser deposition rate (0.5-1Hz). This work led us to the conclusion that the correct interval of deposition parameters is considerably narrower with respect to LAO/STO. The best growth conditions of NGO/STO are similar to those recently proposed for the fabrication of conducting LGO/STO[24] interfaces at high deposition pressures, but with a lower deposition temperature and reduced laser repetition rate. Samples grown at about 700°C, $5\times10^{-2}$ mbar oxygen pressure, laser repetition rate 0.5 Hz, laser fluence on target 1.5 J/cm$^2$, target to sample distance 35 mm are reproducibly metallic. Unlike samples grown at higher temperatures (above 770°C) and lower pressures (below about $5\times10^{-3}$ mbar), their conductance is stable in air for thickness exceeding 4 u.c. and, most importantly, stable against a post-annealing in oxygen. In **Figure 7a**, the RHEED intensity oscillations collected in such conditions are shown. The pattern is indicative of a 2D surface while the intensity oscillations vs time are indicative of a layer-by-layer growth, that is better stabilized, a behavior observed on several samples, after a transient of about 3-4 unit cells. The relaxation of the disorder induced by every single laser shot is clearly seen, as shown in the inset. The RHEED pattern of a 10 u.c. thick film is shown in Figure 7c. We associate the half order rods to a doubling of the interplane distance in the pseudocubic [100] direction, due to the orthorhombic distortion of the NGO unit cell. Such surface reconstruction persists till the end of the deposition, with a 2D pattern demonstrating a high crystallinity.

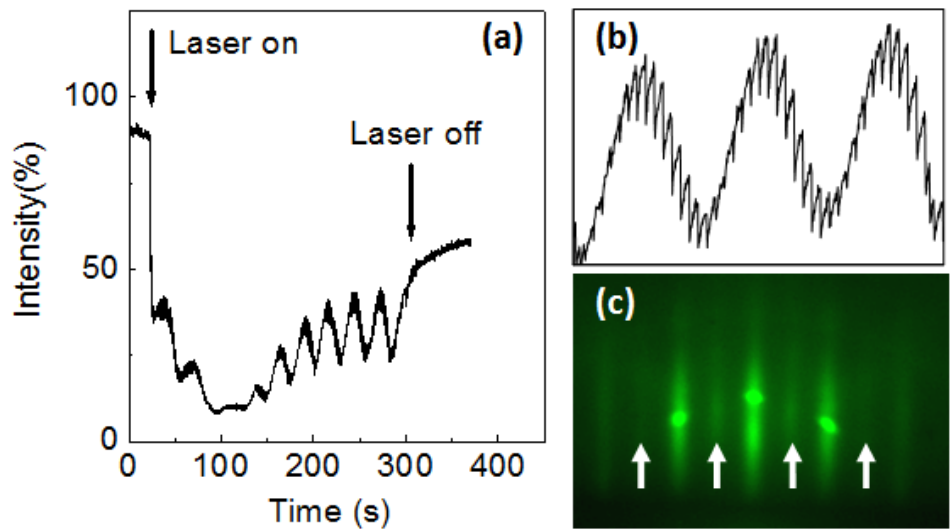

**Figure 7.** a) RHEED oscillations during the growth of a NGO film grown at 700°C and P(O$_2$) = 5·10$^{-2}$ mbar.; b) magnification of a part of Figure 1a; c) final RHEED pattern. Arrows indicate the fractional reflections.

**Acknowledgements**

Financial support by EU under the project OXIDES, by MIUR under Grant Agreement PRIN 2008 - 2DEG FOXI, by European Union Seventh Framework Program (FP7/2007-2013) under grant agreement N. 264098 - MAMA, and by Compagnia di San Paolo is acknowledged. CC and ARL acknowledge funding by the US Department of Energy, Office of Science, Materials Sciences and Engineering Division. Part of this research was supported by ORNL's Shared Research Equipment (ShaRE) User Program, which is sponsored by the Office of Basic Energy Sciences, U.S. Department of Energy.